\documentclass[prd,aps,nofootinbib,showpacs, 10 pt]{revtex4}
\usepackage{amsmath}
\usepackage{amsmath,graphicx,color,epsfig}

\usepackage{ifthen}

\usepackage{epsfig}
\usepackage{amssymb}
\usepackage{graphicx}
\usepackage{amsmath}
\usepackage[center]{subfigure}

\begin{document}
\hfill{}
\title{The Production of ${\it \Xi}_{\rm bb}$ at Photon Collider}

\footnotetext{Supported by the National Natural Science Foundation of China
under Grant Nos 11047109 and 11204067.\\
$^{*}$Corresponding author. Email: yangzj@hpu.edu.cn.}

\author{YANG Zhong-Juan$^{*}$} 
\affiliation{School of Physics and Chemistry, Henan Polytechnic University,
Jiaozuo Henan, 454003, China}
\author{ZHAO Xiao-xia}
\affiliation{School of Physics and Chemistry, Henan Polytechnic University,
Jiaozuo Henan, 454003, China}

\begin{abstract}
The inclusive  production of the doubly heavy baryon ${\it \Xi}_{\rm bb}$
is investigated at polarized and unpolarized  photon collider.
bb pair in both color triplet and sextet have been 
considered to transform into ${\it \Xi}_{\rm bb}$.  The results indicate that
the contribution from color sextet is about 8$\%$ for ${\it \Xi}_{\rm bb}$
production. For the ILC collision energy ranging from $91{\rm GeV}$ to 
$1000{\rm GeV}$, the total cross section of ${\it \Xi}_{\rm bb}$ shows a 
downtrend, i.e., the production of ${\it \Xi}_{\rm bb}$ has the maximal 
rate at $\sqrt {S}=91{\rm GeV}$. Our results 
indicate that the initial beam polarization may be an important asset
for the production of  ${\it \Xi}_{\rm bb}$. At some collision point, 
the production rate of ${\it \Xi}_{\rm bb}$ can be increased 
about 17$\%$ with an appropriate choice of the initial beam polarization.

\end{abstract}

\pacs{13.60.Rj, 12.38.Bx, 14.20.Pt}

\maketitle


The $125 {\rm GeV}$ Higgs-like boson are discovered at ATLAS and CMS, 
which may be a big breakthrough in precise testing of the Standard Model (SM).  
 So the main goal of all kinds of colliders is not only hunting the  
new physics but also testing the SM precisely. In SM, baryon with two 
heavy quarks called doubly heavy baryon, is conformed to exist. 
At present, only the doubly heavy baryon  ${\it \Xi}_{\rm cc}$ has been 
observed by SELEX collaboration 
\cite{Mattson:2002vu,Moinester:2002uw,Ocherashvili:2004hi}. However, the  
decay width and production rate 
 measured at SELEX  are not consistent with 
most of the theoretical predictions. There is no any
evidence for the doubly heavy baryon  ${\it \Xi}_{\rm bb}$ or  
${\it \Xi}_{\rm bc}$  in experiments.
These doubly heavy baryons offer good opportunities to study various 
perturbative or non-perturbative QCD theories, 
especially they can shine on  the color connections of internal QQ pair
and the transformation into the 
color singlet\cite{Han:2006mpa,Jin:2013bra}. 
So it is necessary to study the production 
mechanism more precisely. The theoretical
studies concerning ${\it \Xi}_{\rm bb}$ production can be found in all kinds of 
colliders, such as 
$pp,ep,e^+ e^-$ \cite{Falk:1993gb,Bagan:1994dy,Berezhnoi:1995fy,
Baranov:1995rc,Li:2007vy,Chang:2009va,
Jiang:2012jt,Wang:2012vj,Jiang:2013ej,Yang:2014ita}, etc.. 
These studies show that
the production rate of ${\it \Xi}_{\rm bb}$ is much less than that of
 ${\it \Xi}_{\rm cc}$.  Thus
a more cleaning environment is needed for investigating the production 
of ${\it \Xi}_{\rm bb}$. The 
International Linear Collider (ILC) may be a good platform for studying
the production of ${\it \Xi}_{\rm bb}$.

The International Linear Collider (ILC) can play a key role in the 
precise measurements of future elementary particle physics. The collider 
has many advantages.
First, it is possible to run at arbitrary center of mass (CM) energies between
 $91 {\rm GeV}$ to $1000 {\rm GeV}$ with high luminosity\cite{Baer:2013cma}. 
Second, the ILC can provide an environment in which high energy collisions 
can be measured with high precision and one can perform physics analyses
on all final states of the  decay particles. 
The last, at the ILC, spin effects can provide us crucial new handles on 
investigation of all kinds of physics, which can be
realized by easily controlling of the polarization of the initial beams. 
 The photon photon collider can be realized at ILC by Compton 
backscattering laser light\cite{Ginzburg:1982yr}, which  may produce very high energy photons
, and photon photon collider has all of the above 
advantages at ILC. Thus many physics programs can be employed at the 
photon collider, during which the production of  doubly heavy baryon 
${\it \Xi}_{\rm bb}$ 
is a potential one. The advantages of ILC may be helpful for looking for the 
doubly heavy baryon ${\it \Xi}_{\rm bb}$. 

The factorization can be used to handle the production of 
 ${\it \Xi}_{\rm bb}$ \cite{Bodwin:1994jh,Ma:2003zk}. The 
first is the production of heavy quark pair bb, which can be calculated using
perturbative QCD. 
 The second is 
the non-perturbative transformation of bb into  ${\it \Xi}_{\rm bb}$, 
which can be handled with non-relativistic QCD (NRQCD) because of the 
small velocity of b quark in the rest frame of   ${\it \Xi}_{\rm bb}$
\cite{Bodwin:1994jh,Ma:2003zk}.
In this letter, we systematically
investigate  the inclusive production of
 ${\it \Xi}_{\rm bb}$ at photon photon collider. It is found that 
proper choice of the initial beam  polarization may increase  
the production rate of ${\it \Xi}_{\rm bb}$
up to 17$\%$.
We hope this can further improve the theoretical predictions and be helpful 
for the research of ${\it \Xi}_{\rm bb}$ at ILC.


For the leading order contribution at photon collider, the  production 
of doubly heavy baryon ${\it \Xi}_{\rm bb}$ can be described via the 
following inclusive process
\begin{equation}
\label{wen1}
\gamma(p_1,\lambda_1)+\gamma(p_2,\lambda_2) \to \Xi_{\rm bb}(k)+\bar{b}(p_3)
+\bar{b}(p_4)+X_{\rm N} ,
\end{equation}
where $\lambda_1$ and $\lambda_2$ are the polarizations of the photons,
the momenta of the corresponding particles are respectively denoted as
 $p_i(i=1,2,3,4)$ and $k$.  $X_{\rm N}$ is the non-perturbative unobserved state. 
Here, the high energy photons can be provided by Compton backscattered 
laser beam \cite{Ginzburg:1982yr} and the corresponding process is
shown in Figure 1. The total effective cross section 
for ${\it \Xi}_{\rm bb}$  production can be written as
\begin{equation}
\label{eeq14}
d\sigma (S)=\int_{0}^{y_{max}} dy_1 \int_{0}^{y_{max}} dy_2 f_{\gamma}^e
(y_1,P_{\rm e},P_{\rm L}) f_{\gamma}^e(y_2,P_{\rm e},P_{\rm L}) d\hat{\sigma}
(\hat s,\lambda_1,\lambda_2),
\end{equation}
where $S$ is the square of $e^+e^- $ center of mass energy, $\hat s=(p_1+p_2)^2$, 
$P_{\rm L}(P_{\rm e})$ is the polarization of the initial laser (electron) beam.
$f_{\gamma}^e(y,P_{\rm e},P_{\rm L})$ is the  photons'  energy spectrum, and it's 
 normalized formation is   
\begin{equation}
\label{rre}
f_{\gamma}^e(y,P_{\rm e},P_{\rm L})={\cal N}^{-1}[\frac{1}{1-y}-y+(2r-1)^2
-P_{\rm e}P_{\rm L}xr(2r-1)(2-y)],
\end{equation}
$\cal N$ is the normalization factor,
$r=y/(x-xy)$. $y$ is the fraction of the photon energy
obtained from the electron in center of mass system. Here, we have 
$0\leq y \leq x/x+1$ and $ x=4 E_{\rm L} E_{\rm e}/m_e^2$. The maximum of x 
can only reach $2(1+\sqrt{2})$, which can avoid the creation of $e^+e^-$ 
pairs from the initial laser light and the backscattered laser beam.


\begin{figure}
\centering
\includegraphics[width=0.4\textwidth]{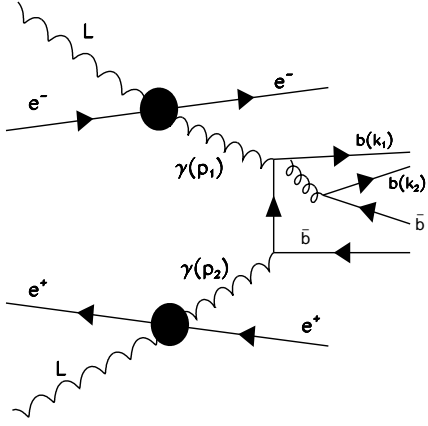}
\caption{The perturbative production of bb pair in two photon process which 
resulting from Compton backscattering laser.}
\label{xdew}
\end{figure}

\begin{figure}
\centering
\includegraphics[width=0.5\textwidth]{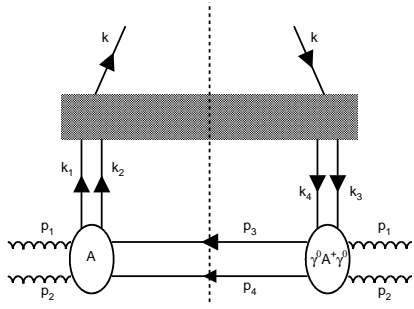}
\caption{The contributions in eq.(4)}
\label{xdew}
\end{figure}

The differential cross section $ d\hat{\sigma}(\hat s,\lambda_1,\lambda_2)$ 
in  Eq.(1) can be expressed as
\begin{eqnarray}
\label{wen3}
&&d\hat{\sigma}(\hat s,\lambda_1,\lambda_2)=\frac{1}{4}\frac{1}
{2\hat s} \sum_{X_{\rm N}}
\frac{d^3 \bf k}{(2\pi)^3} \int  \frac{d^3p_3}{(2\pi)^3 2E_3} \frac{d^3p_4}
{(2\pi)^3 2E_4}(2\pi)^4 \delta^4(p_1+p_2-k-p_3-p_4-P_{X_{\rm N}}) \nonumber \\
&&\cdot  \int \frac{d^4k_1}{(2\pi)^4} \frac{d^4k_3}{(2\pi)^4}
\frac{1}{2}A_{ij}(k_1,k_2,p_3,p_4,\lambda_1,\lambda_2)
\frac{1}{2} (\gamma^0 A^\dagger(k_3,k_4,p_3,p_4,\lambda_1,\lambda_2)\gamma^0)_{kl}
\nonumber \\
&& \cdot \int d^4x_1 d^4x_3 e^{-ik_1 \cdot x_1+ik_3 \cdot x_3} \langle
0|b_k(0)b_l(x_3)|{\it \Xi}_{\rm bb}+X_{\rm N} 
\rangle \langle {\it \Xi}_{\rm bb}+X_{\rm N}|\bar{b_i}(x_1)
\bar{b_j}(0)|0 \rangle, 
\end{eqnarray}
with the photons' polarization
\begin{eqnarray}
\lambda_i&=&\frac{1}{f_{\gamma}^e(y_i,P_{\rm e}^{(i)},P_{\rm L}^{(i)})\cal N}
\{xrP_{\rm e}^{(i)}[1+(1-y)(2r-1)^2]-(2r-1)P_{\rm L}^{(i)}
[\frac{1}{1-y}+1-y]\},  i=1,2.
\end{eqnarray}
Here we imply the average over the photons' polarization and the
summation over the spin and color of all final state particles.
$ A_{ij}(k_1,k_2,p_3,p_4,\lambda_1,\lambda_2)$ is the scattering amplitude 
for the production of heavy quark pair bb in photon photon collision, 
which can be calculated using 
perturbative QCD. $k_1$, $k_2$ denote the four-momentum of the
internal bb quarks and $k_1=k_2=k/2$.
$b(x)$ is the Dirac field for b quark. 
Fig.\,2 shows the contributions in Eq.\,(4).
Employing the non-relativistic 
normalization for ${\it \Xi}_{\rm bb}$ and the translation
invariance to handle the sum over $X_{\rm N}$, one can obtain
\begin{eqnarray}
\label {wen6}
d\hat{\sigma}(\hat s,\lambda_1,\lambda_2 )&=&\frac{1}{4}\frac{1}
{2\hat{s}}\frac{d^3\bf k}
{(2\pi)^3}\int \frac{d^3p_3}
{(2\pi)^3 2E_3}\frac{d^3p_4}{(2\pi)^3 2E_4}\int \frac{d^4k_1}{(2\pi)^4}
\frac{d^4k_3}{(2\pi)^4}\frac{1}{2}
 A_{ij}(k_1,k_2,p_3,p_4,\lambda_1,\lambda_2)\nonumber\\
&&\cdot \frac{1}{2} (\gamma^0 A^\dagger(k_3,k_4,p_3,p_4,\lambda_1,\lambda_2)
\gamma^0)_{kl} \cdot\frac{1}
{v^0} (2\pi)^4 \delta^4(k_1-m_{\rm b} v)(2\pi)^4
\delta^4(k_2-m_{\rm b}v)(2\pi)^4
\delta^4(k_3-m_{\rm b}v)\nonumber \\
&&\cdot [-{(\delta_{a_1a_4}\delta_{a_2a_3}+\delta_{a_1a_3}\delta_{a_2a_4})
({P_v}^T C \gamma_5 P_v)_{ji}(P_v \gamma_5 C
{P_v}^T)_{lk}}
\cdot{h_1} \nonumber \\
&&+(\delta_{a_1 a_4} \delta_{a_2 a_3} - \delta_{a_1 a_3} \delta_{a_2a_4})
({P_v}^T C \gamma^{\mu} P_v)_{ji}(P_v \gamma^{\nu}
C {P_v}^T)_{lk}
(v_{\mu}v_{\nu}-g_{\mu\nu})\cdot {h_3} ],
\end{eqnarray}
where $P_v=1+{\gamma}\cdot v/2$, $C=i\gamma^2 \gamma^0$,  
and $v^{\mu}=k^{\mu}/M_{{\it \Xi}_{\rm bb}}$. The color and Dirac indices are
 respectively 
denoted as $a_i(i=1, 2, 3, 4)$ and $i$, $j$, $k$, $l$. For bb quark system, 
there are two states contributing to the production of ${\it \Xi}_{\rm bb}$,
 one  is 
bb  in $^3S_1$ and color triplet, the other is in $^1S_0$ and color sextet.  
The corresponding probability to transform into 
the baryon can be respectively represented by $ h_1$ and $ h_3$,
which should be determined by
non-perturbative QCD. Under NRQCD, as noted in \cite{Bagan:1994dy,Ma:2003zk}, 
one can have $h_1\approx h_3=|{\it \Psi}_{\rm bb}(0)|^2$.


In principle, ILC can run on any energy stage from $91{\rm GeV}$ to 
$1000{\rm GeV}$ with
high luminosity. So we choose different CM energies for the numerical
calculation of ${\it \Xi}_{\rm bb}$. 
Here we take  $\alpha_s(m_Z)=0.1185$ and 
let the leading order $\alpha_s$ run. With
the renormalization scale being chosen to be $2 m_{\rm b}$, one can have
$\alpha_s(2 m_{\rm b})=0.178$ for $m_{\rm b}=5.1{\rm GeV}$.

The total effective cross sections for  
${\it \Xi}_{\rm bb}$ at $\sqrt {S}=500{\rm GeV}$ are given in Table 1.
In Ref.\cite{Bagan:1994dy}, $|{\it \Psi}_{\rm bb}(0)|^2=0.152{\rm GeV^3}$.
Here, the values of $|{\it \Psi}_{\rm bb}(0)|^2$  change from 
$0.136 {\rm GeV^3}$ to $0.168 {\rm GeV^3}$. One can 
find that, when the parameter $h_1$($h_3$) is raised(reduced) 10$\%$, 
the total
cross section correspondingly increase (drop down) about 10$\%$ . The exact 
values of
 $h_1$ and $h_3$ have impacts on the production rate of ${\it \Xi}_{\rm bb}$. 
For the production of ${\it \Xi}_{\rm bb}$, the
contribution from bb pair in the color sextet is approximately 10$\%$
of that from the color triplet one. 
The total cross section promptly drops down when b quark mass changes from 
$m_{\rm b}=4.1{\rm GeV}$ to $m_{\rm b}=5.35 {\rm GeV}$, i.e., the production rate 
of ${\it \Xi}_{\rm bb}$ reduces by  a
factor of 67$\%$ corresponding to the 30$\%$ enlargement of b quark mass.

Employing the  polarizations of the electron (laser) beams, we calculate
the  total effective cross section of ${\it \Xi}_{\rm bb}$ for
  $\sqrt {S}=250{\rm GeV}$
at photon collider. The results are shown in 
Table 2. One can notice that the choice of 
$(P_{\rm e1},P_{\rm e2};P_{\rm L1},P_{\rm L2})=(0.85,-0.85;+1,-1)$
can increase the production rate of ${\it \Xi}_{\rm bb}$ by approximately 17$\%$, 
which means that the polarization of initial beams is an 
important asset for ${\it \Xi}_{\rm bb}$ production.
Compared with the results in Table 1, one can find that 
the  cross section of
 ${\it \Xi}_{\rm bb}$  at $\sqrt {S}=250{\rm GeV}$ is twice larger than that 
at $500{\rm GeV}$ CM energy.
 
\vskip 5mm
\noindent{\footnotesize {\bf Table 1.} Total cross section (in unit: fb)
of ${\it \Xi}_{\rm bb}$ production at $\sqrt {S}=500{\rm GeV}$.

\vskip 2mm \tabcolsep 3pt
\centerline{\footnotesize
\begin{tabular}{ccccccc}
\hline
$(h_1, h_3)$\,GeV$^3$& $m_{\rm b}$=4.1\,GeV& $m_{\rm b}$=4.35\,GeV&
$m_{\rm b}$=4.6\,GeV& $m_{\rm b}$=4.85\,GeV&$m_{\rm b}$=5.1\,GeV
&$m_{\rm b}$=5.35\,GeV\\
\hline
 (0.136,0) & $7.13\times 10^{-3}$ & $5.45\times 10^{-3}$ 
& $4.26\times 10^{-3}$  & $3.37 \times 10^{-3} $ & $2.66 \times 10^{-3} $ 
& $2.16 \times 10^{-3} $ \\ 
(0,0.136)   &$8.08 \times 10^{-2} $ & $6.28 \times 10^{-2} $
 &$4.99 \times 10^{-2} $  &$4.01 \times 10^{-2} $ &$3.20 \times 10^{-2} $
&$2.65 \times 10^{-2} $ \\ 
(0.136,0.136) &$8.79 \times 10^{-2} $ &$6.83 \times 10^{-2} $
 &$5.42 \times 10^{-2} $ &$4.35 \times 10^{-2} $& $3.47 \times 10^{-2} $
 &$2.87 \times 10^{-2} $ \\  
(0.144,0) & $7.55 \times 10^{-3} $ & $5.77 \times 10^{-3} $
  & $4.51 \times 10^{-3}  $ & $ 3.57 \times 10^{-3} $ & $2.82 \times 10^{-3} $
  & $2.29 \times 10^{-3} $ \\  
 (0,0.144)  &$8.54 \times 10^{-2} $&$6.64 \times 10^{-2} $ 
 &$5.28 \times 10^{-2} $ &$4.25 \times 10^{-2} $ & $3.39 \times 10^{-2} $
&$2.81 \times 10^{-2}$ \\ 
(0.144,0.144) &$9.32 \times 10^{-2}$ &$7.22 \times 10^{-2}$ 
&$5.73 \times 10^{-2}$ &$4.61 \times 10^{-2}$& $3.67 \times 10^{-2}$
 &$3.04 \times 10^{-2}$ \\ 
 (0.152,0) & $7.97 \times 10^{-3} $ & $6.09 \times 10^{-3} $ 
& $4.77 \times 10^{-3}  $  & $3.77 \times 10^{-3} $ & $2.97 \times 10^{-3} $ 
& $2.42 \times 10^{-3} $ \\ 
 (0,0.152) &$9.02 \times 10^{-2}$& $7.01 \times 10^{-2}$ 
 &$5.58 \times 10^{-2}$ &$4.47 \times 10^{-2}$  & $3.58 \times 10^{-2}$
&$2.96\times 10^{-2}$ \\  
(0.152,0.152) &$9.82 \times 10^{-2}$ &$7.62 \times 10^{-2}$
&$6.06 \times 10^{-2}$ &$4.85 \times 10^{-2}$ & $3.88 \times 10^{-2}$
 &$3.20 \times 10^{-2}$ \\  
(0.160,0) & $8.39 \times 10^{-3} $ & $ 6.42 \times 10^{-3} $
 & $5.02 \times 10^{-3} $ & $3.96 \times 10^{-3} $ & $ 3.12 \times 10^{-3 }  $
& $2.55 \times 10^{-3} $ \\  
 (0,0.160)  &$9.5 \times 10^{-2}$&$7.38 \times 10^{-2}$  
&$5.86 \times 10^{-2}$  & $4.71 \times 10^{-2}$ &$3.77 \times 10^{-2}$
  &$3.11 \times 10^{-2}$\\  
 (0.160,0.160) &$10.34 \times 10^{-2}$ &$8.02 \times 10^{-2}$
&$6.36 \times 10^{-2}$ &$5.11 \times 10^{-2}$ &$4.08 \times 10^{-2}$ 
&$3.37 \times 10^{-2}$ \\  
(0.168,0) & $8.81\times 10^{-3} $& $6.74\times 10^{-3} $
 & $5.27\times 10^{-3}  $  & $4.16\times 10^{-3} $ & $ 3.29\times 10^{-3} $
& $2.67\times 10^{-3} $ \\  
(0,0.168) &$9.98 \times 10^{-2}$& $7.74 \times 10^{-2}$  
&$6.16 \times 10^{-2}$ &$4.95 \times 10^{-2}$  & $3.96 \times 10^{-2}$ 
&$3.28 \times 10^{-2}$ \\  
(0.168,0.168) &$10.86 \times 10^{-2}$&$8.41 \times 10^{-2}$  
&$6.69 \times 10^{-2}$ &$5.37 \times 10^{-2}$ & $4.29 \times 10^{-2}$
&$3.55 \times 10^{-2}$\\ 
\hline 
\end{tabular}}}

\vskip 0.5\baselineskip
\vskip 2mm

\noindent{\footnotesize {\bf Table 2.} Results for the effective cross 
section  (in unit: fb) of ${\it \Xi}_{\rm bb}$ (for $m_{\rm b}=5.1{\rm GeV}$ )
at $\sqrt {S}=250 {\rm GeV}$.

\vskip 2mm \tabcolsep 3pt
\centerline{\footnotesize
\begin{tabular}{cccc}
\hline
 $h_1(\rm GeV^3)$  & $0$     &$0.152$ &$0.152$ \\ 
  $h_3(\rm GeV^3)$ & $0.152$ &$0$     &$0.152$ \\ 
\hline
spin averaged    &$7.89 \times 10^{-2}$ & $5.01\times 10^{-3}$
&$8.39 \times 10^{-2}$\\
$(P_{e1},P_{e2};P_{L1},P_{L2})=(0.85,0.85;-1,-1)$& $6.94 \times 10^{-2}$ 
&$4.42 \times 10^{-3}$   & $7.38 \times 10^{-2}$\\ 
$(P_{e1},P_{e2};P_{L1},P_{L2})=(0.85,0.85;+1,+1)$ &$7.57 \times 10^{-2}$ 
&$7.46 \times 10^{-3}$   & $8.32 \times 10^{-2}$\\ 
$(P_{e1},P_{e2};P_{L1},P_{L2})=(0.85,-0.85;-1,+1)$ &$7.50 \times 10^{-2}$ 
&$4.88 \times 10^{-3}$   & $7.99 \times 10^{-2}$\\ 
$(P_{e1},P_{e2};P_{L1},P_{L2})=(0.85,-0.85;+1,-1)$  & $9.48 \times 10^{-2}$  
& $3.24 \times 10^{-3}$  & $9.8 \times 10^{-2}$\\ \hline
\end{tabular}}}

\vskip 1\baselineskip

In the following numerical calculation, we take the parameters as that 
in Ref.\cite{Bagan:1994dy}
\begin{eqnarray}
&&|{\it \Psi}_{\rm bb}(0)|^2={\rm 0.152GeV^3}, \hspace{2.7cm}
m_{\rm b}={\rm 5.1 GeV}.
\end{eqnarray}

\begin{figure}
\centering
\includegraphics[width=0.33\textwidth]{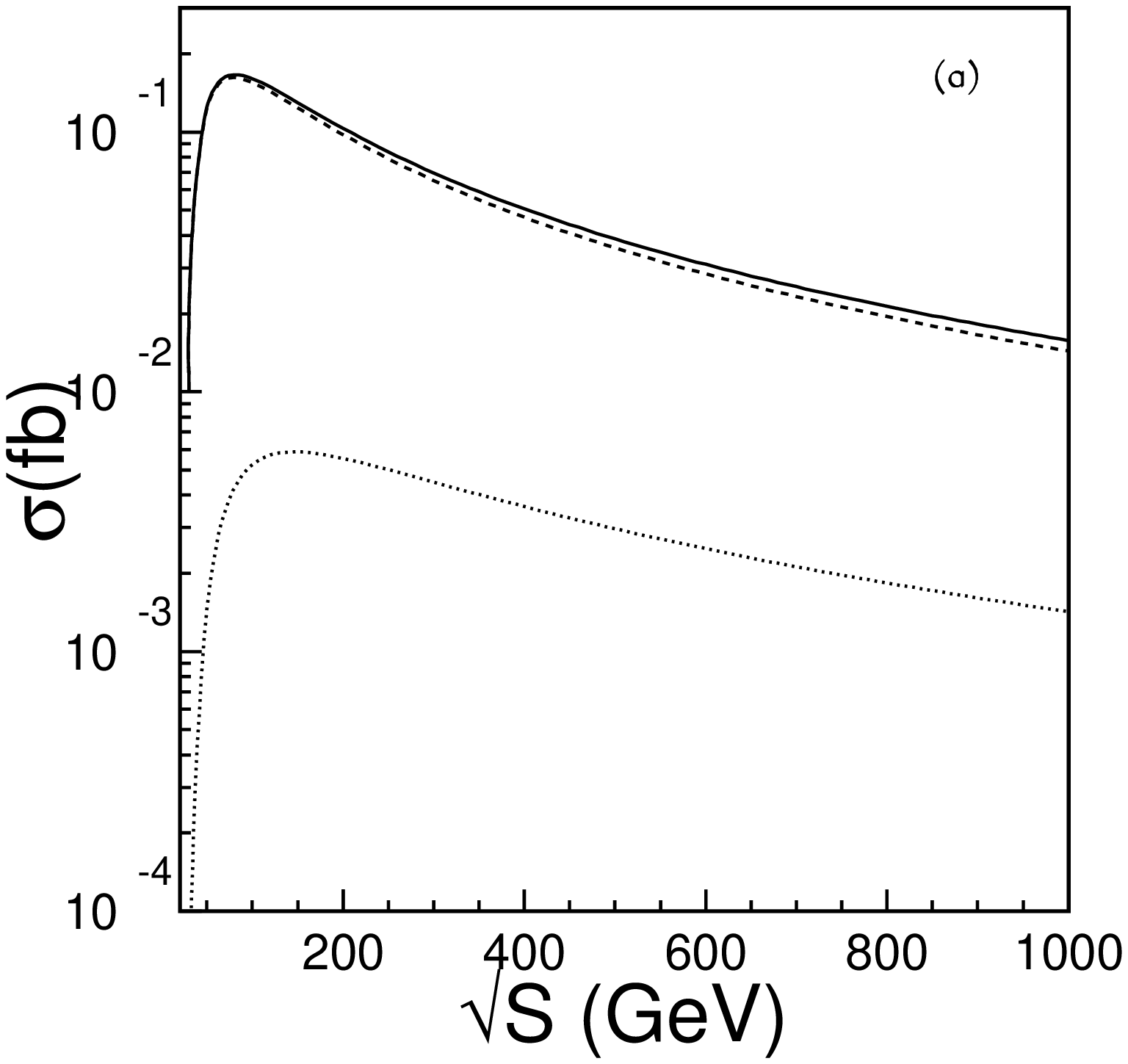}
\includegraphics[width=0.33\textwidth]{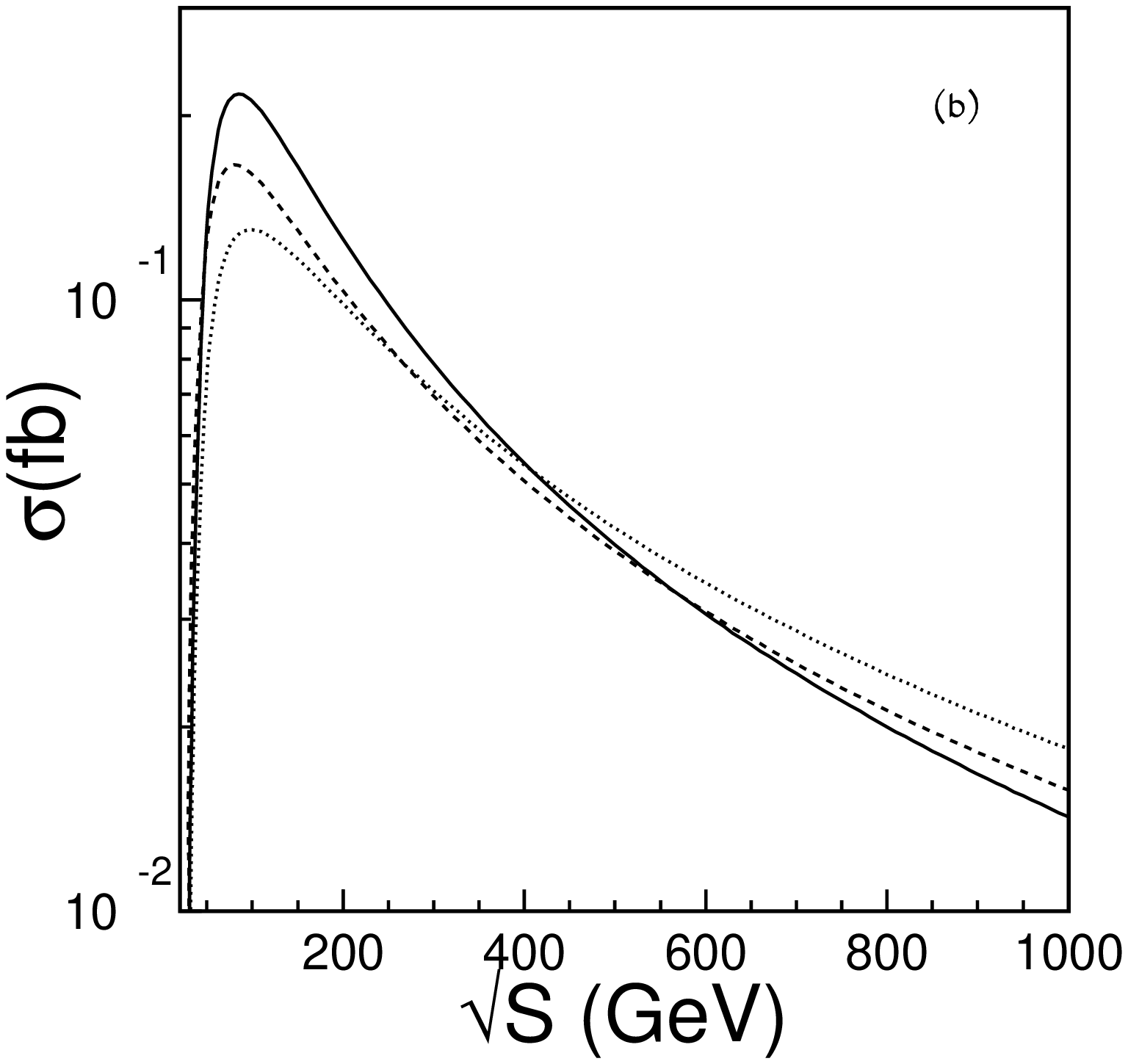}
\caption{The total cross sections of ${\it \Xi}_{\rm bb}$  versus the 
center of mass energy. (a)The solid line stands for
$h_3=h_1=|{\it \Psi}_{\rm bb}(0)|^2$, the dotted stands for 
$h_1=|{\it \Psi}_{\rm bb}(0)|^2$
and $h_3=0$, and the dashed stands for
$h_3=|{\it \Psi}_{\rm bb}(0)|^2$ and $h_1=0$. (b)The solid line stands for
 $(P_{\rm e1},P_{\rm e2};P_{\rm L1},P_{\rm L2})=(0.85,-0.85;+1,-1)$, 
the dotted stands for $(P_{\rm e1},P_{\rm e2};P_{\rm L1},P_{\rm L2})=(0.85,0.85;+1,+1)$,
and the dashed line stands for spin averaged .}
\label{energyto}
\end{figure}

The total cross sections of ${\it \Xi}_{\rm bb}$
versus the collision energy are shown in Fig.\ref{energyto}.
bb pair in different color states are calculated in Fig.\ref{energyto}(a).
The contribution from color triplet is much larger than color
sextet. One can find that the largest production rate appears around
$\sqrt {S}=91{\rm GeV}$, which is the  supposed energy at Giga-Z program. 
The cross sections of ${\it \Xi}_{\rm bb}$ with all kinds of 
polarizations of the initial beams are considered. 
The choices of $(P_{\rm e1},P_{\rm e2};P_{\rm L1},P_{\rm L2})=(0.85,0.85;+1,+1)$
and  $(P_{\rm e1},P_{\rm e2};P_{\rm L1},P_{\rm L2})=(0.85,-0.85;+1,-1)$ have 
large impacts on
the production rate, which are shown  in Fig.\ref{energyto}(b).
 For different CM energy, the choice of initial beam 
polarization shows different effect. When the CM 
energy is less than $400 {\rm GeV}$,  the choice of 
 $(P_{\rm e1},P_{\rm e2};P_{\rm L1},P_{\rm L2})=(0.85,-0.85;+1,-1)$ can increase 
the production 
rate of ${\it \Xi}_{\rm bb}$, while  $(P_{\rm e1},P_{\rm e2};P_{\rm L1},P_{
rm L2})=(0.85,0.85;1,1)$ has 
the same impact when $\sqrt {S} > 400 {\rm GeV}$.

\begin{figure}
\centering
\includegraphics[width=0.31\textwidth]{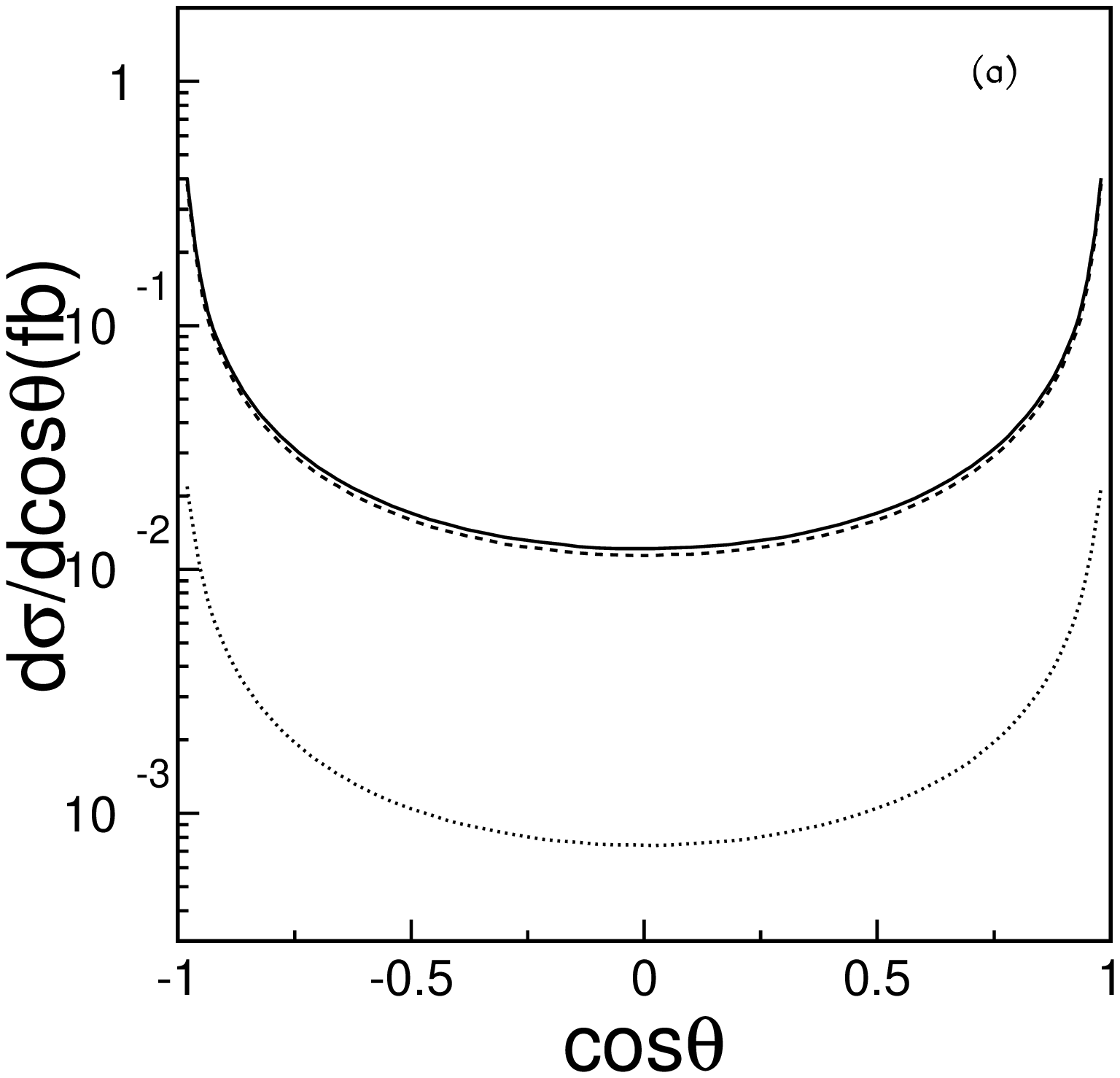}
\includegraphics[width=0.31\textwidth]{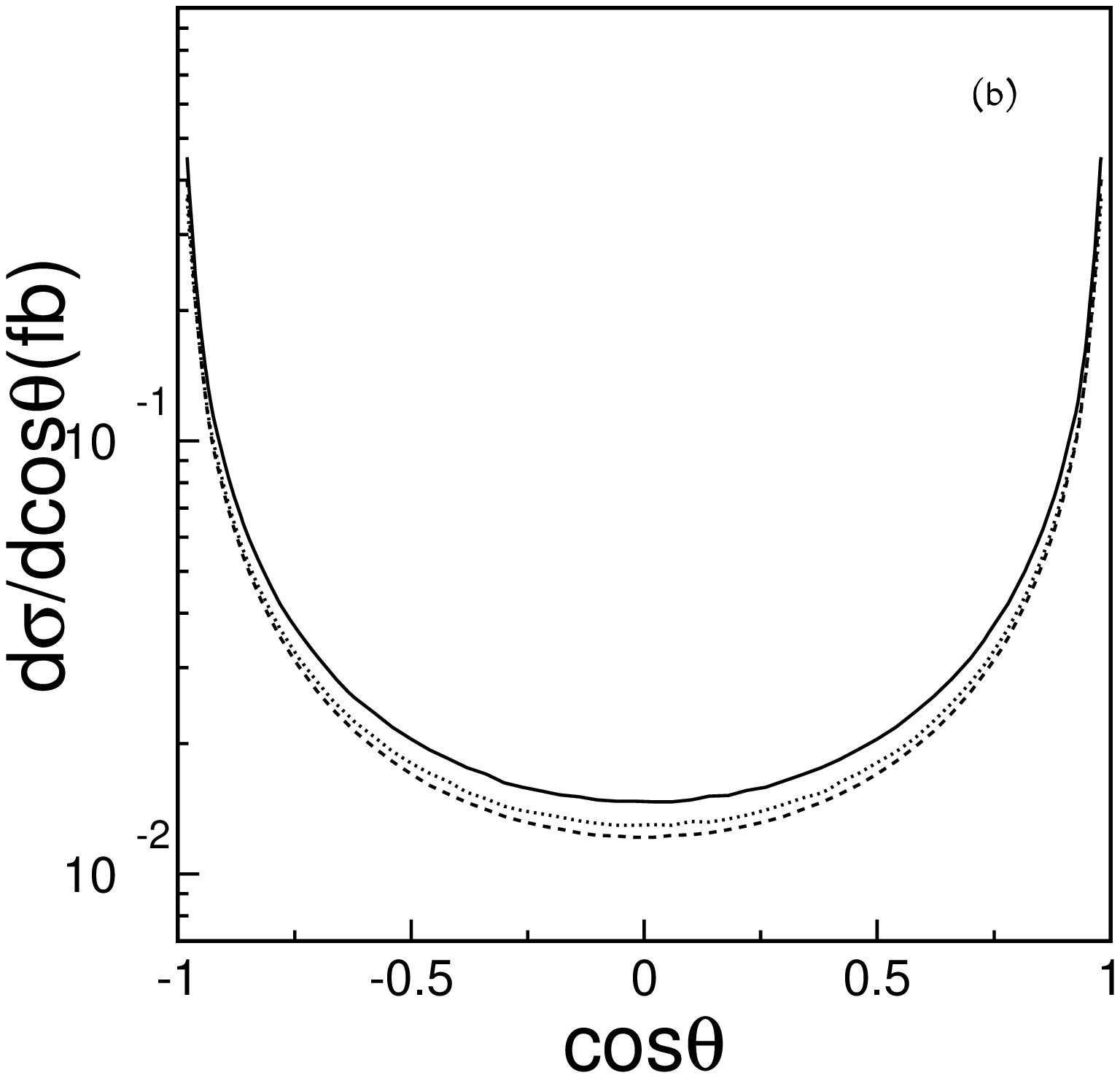}
\caption{The distributions of $\cos\theta$. (a) The solid line stands for
$h_3=h_1=|{\it \Psi}_{\rm bb}(0)|^2$, the dotted stands for 
$h_1=|{\it \Psi}_{\rm bb}(0)|^2$
and $h_3=0$, and the dashed stands for
$h_3=|{\it \Psi}_{\rm bb}(0)|^2$ and $h_1=0$. (b)The solid line stands 
for $(P_{\rm e1},P_{\rm e2};P_{\rm L1},P_{\rm L2})=(0.85,-0.85;+1,-1)$, 
the dotted stands for $(P_{\rm e1},P_{\rm e2};P_{\rm L1},P_{\rm L2})=(0.85,0.85;-1,-1)$,
and the  dashed line stands for spin averaged.}
\label{efwtg}
\end{figure}

The $\cos\theta$-distributions, $x_{\rm T}$-distributions and $x$-distributions 
 for ${\it \Xi}_{\rm bb}$ are given in 
Fig.\ref{efwtg}, \ref {xdew} and \ref{dhfgk} respectively. 
Here the collision energy is taken
to be $250 {\rm GeV}$. $\theta$ is the angle between the initial $e^+$ beam and 
the final ${\it \Xi}_{\rm bb}$ moving direction. $x_{\rm T}=2p_{\rm T}/ \sqrt{S}$
and $x=2E/ \sqrt{S}$, with $p_{\rm T}$ and $E$ 
 the transverse momentum and energy of ${\it \Xi}_{\rm bb}$.
The contributions of bb pair in 
both color states have the same decency in each  
distribution for ${\it \Xi}_{\rm bb}$ production at $ \sqrt{S}=250 GeV$. 
Different polarizations of the initial beams slightly change the distributions, 
which can be seen 
from Fig.\ref{efwtg}(b),\ref {xdew}(b) and  \ref{dhfgk}(b). 

For bb quark system, neglecting the relative orbital angular momentum 
and considering anti-symmetry property of the wave function, there are 
only two states contributing to the production of ${\it \Xi}_{\rm bb}$. 
The method can be used to study the production of doubly heavy baryon 
${\it \Xi}_{\rm bc}$. But for ${\it \Xi}_{\rm bc}$ production,
one have more hadronic matrix elements to describe the internal bc quark
system, which is due to no restriction from Pauli principle for the heavy
quark pair bc.


In summary, we investigate the production of the doubly heavy
baryon ${\it \Xi}_{\rm bb}$ at $\gamma \gamma$ collider. We find that for
 ${\it \Xi}_{\rm bb}$ 
production, the contribution from the color sextet is about 8$\%$. Therefore, 
the  color triplet and  sextet should both be included for the 
production of ${\it \Xi}_{\rm bb}$.
The value of $m_{\rm b}$ has large impact on the  ${\it \Xi}_{\rm bb}$ production.
For a fixed value of  $m_{\rm b}$ ,  the largest production rate appears around
$\sqrt {S}=91{\rm GeV}$, as the collision energy running up, the total 
cross section drops down. 
For different collision energy, there are different choices of the initial 
beam polarization which can increase the production rate of ${\it \Xi}_{\rm bb}$. 
For example, The production rate of ${\it \Xi}_{\rm bb}$ can be increased about 
17$\% $
with the initial beam polarization  
$(P_{\rm e1},P_{\rm e2};P_{\rm L1},P_{\rm L2})$ $=(0.85,-0.85;+1,-1)$ at  
$\sqrt {S} =250 {\rm GeV}$. The enhancement is 
larger than the contribution from  the color sextet. These results 
indicate that the initial beam polarization may be an important asset
for the production of  ${\it \Xi}_{\rm bb}$. 
We hope these results can be helpful for the
investigation of  ${\it \Xi}_{\rm bb}$  production at ILC.


\begin{figure}
\centering
\includegraphics[width=0.33\textwidth]{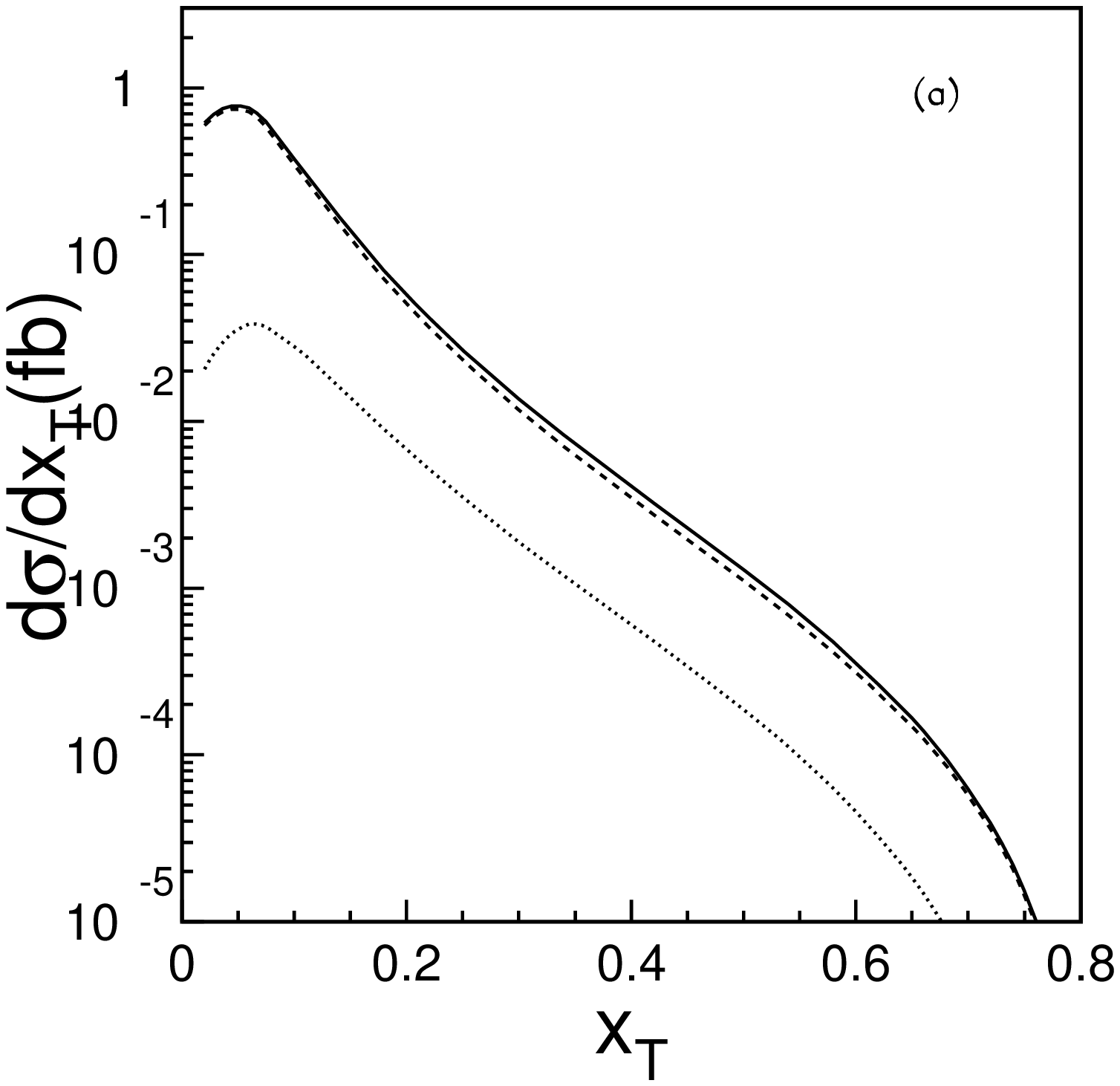}
\includegraphics[width=0.33\textwidth]{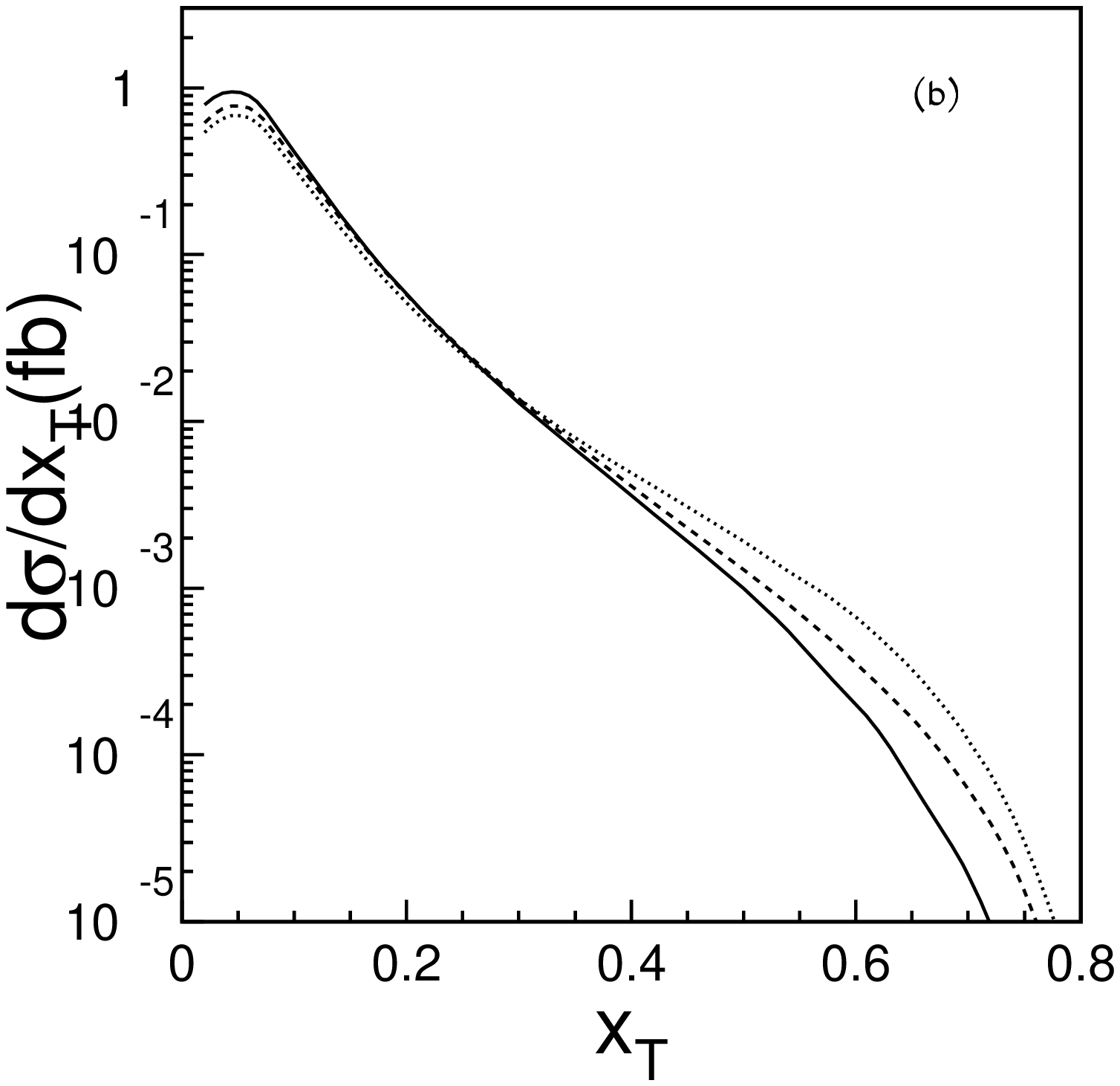}
\caption{Same as Fig.\ref{efwtg}, but for $x_{\rm T}$-distributions.}
\label{xdew}
\end{figure}
\begin{figure}
\centering
\includegraphics[width=0.33\textwidth]{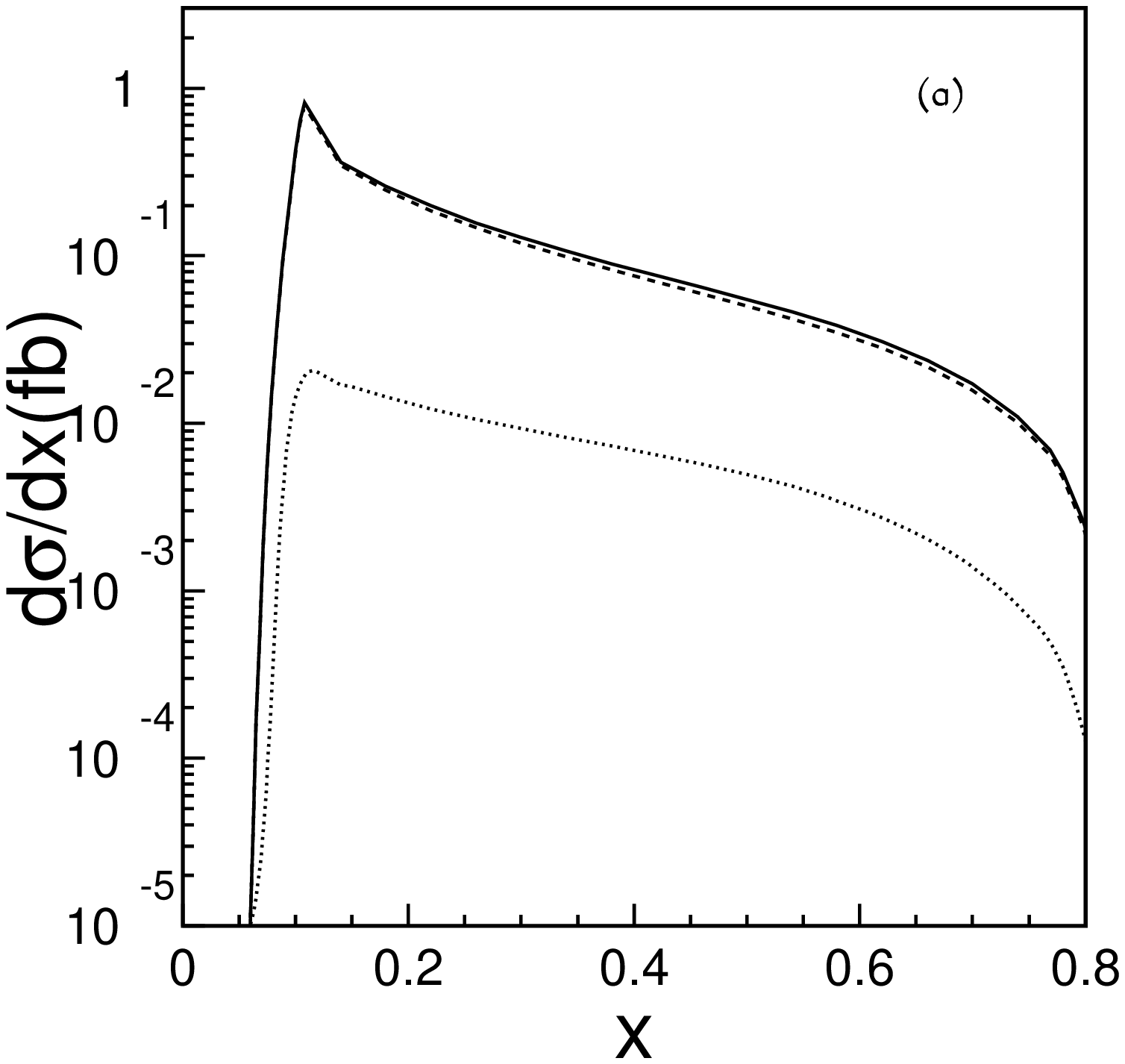}
\includegraphics[width=0.33\textwidth]{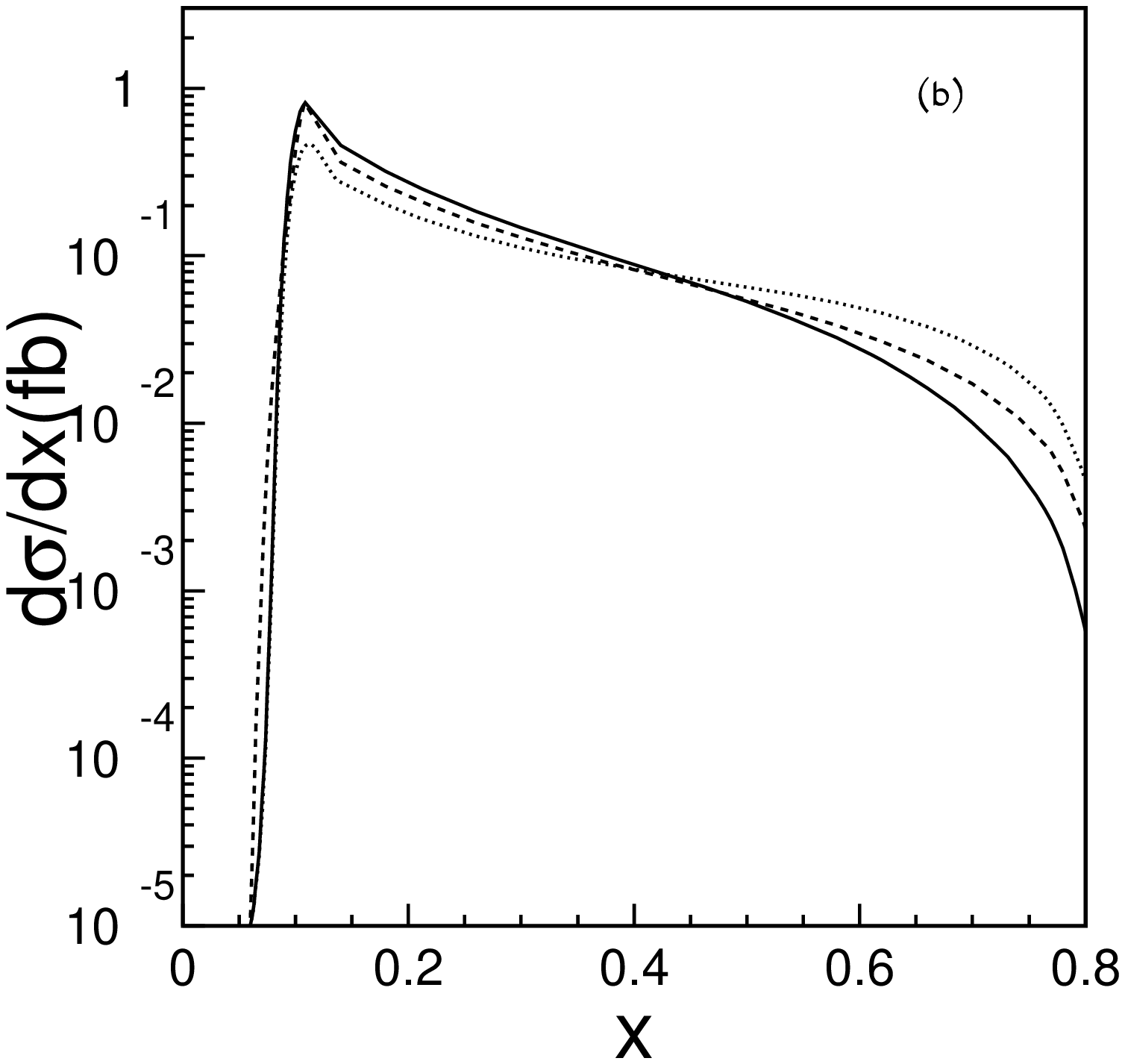}
\caption{Same as Fig.\ref{efwtg}, but for $x$-distributions.}
\label{dhfgk}
\end{figure}

\end{document}